\documentstyle[prd,aps,epsfig,floats]{revtex}
\tighten
\begin{document}
\draft

\title
{Electron Screening in ${^7Be} + p \longrightarrow {^8B}+\gamma$ reaction}

\author{V.~B.~Belyaev${}^{1,2}$ \and D.~E.~Monakhov${}^1$ \and
D.~V.~Naumov${}^{1,3}$ \and and F.~M.~Pen'kov${}^1$}
\address{${}^1$ Joint Institute for Nuclear Research, Dubna, Russia \\
         ${}^2$ Research Center for Nuclear Physics, Osaka University,
         Japan \\
         ${}^3$ Physics Department, Irkutsk State University, Irkutsk
         664003, Russia \\
         }
\date{\today}
\maketitle

\begin{abstract}
We evaluate the effect of screening by bound electron in 
${(^7Be,e)} + p \longrightarrow {(^8B,e)} +\gamma$ transition in the framework
of the adiabatic representation of the three particle problem.
Comparison with two approaches (united nucleus and static) is presented.
We discuss possible applications of this effect both for Solar Neutrinos and 
low energy fusion experiments. 
\end{abstract}
\section{Introduction}

In the recent years an increasing interest has been devoted to the accurate
estimation of electron screening effects both for the solar plasma fusion 
rates and low energy Earth experiments 
(see Ref.~\cite{Gruzinov,Shoppa,Salpeter,Mitler,Carraro,Brown,Langanke} 
and references therein). 

As a rule, one considers electrostatic screening in the solar plasma. This
approach being classical or quantum correctly reflects the major properties
of a process only for high relative velocity of the colliding nuclei, 
when electron density is supposed to be unchanged during the collision.

In the case, when relative velocity of the nuclei is too small in comparison
with the electron one, the electron density is changed accordingly to any
relative configuration of the nuclei. Thus the electrons have an impact on a
kinetic energy shift of the nuclei. Such a process is considered in the
adiabatic approach, which comes from the well known Born-Oppenheimer method.
 
An accurate treatment in the adiabatic approach of electrons from the 
continuum spectrum requires an additional research, while screening
effects by bound electrons could be easily considered in the framework of the
adiabatic approach (see, for example Ref.~\cite{Melezhik}).

For not very dense stars like the Sun, the commonly accepted Debye-H\"uckel 
approximation on the calculation of screening effect is not quite adequate.
Recently this question was discussed by A.~Dar and G.~Shaviv \cite{Dar}.
This fact is usually of no importance
considering that the screening due to the plasma electrons is by itself
rather small (see Ref.~\cite{Gruzinov,Dar}. On the other hand, the low-lying
bound electrons on a nucleus do screen the electric charge of the colliding 
nuclei much effectively. Not only the stellar plasma but Earth 
experiments have such a feature. 

Indeed, laboratory experiments are performed with atomic or molecular
targets, and the ionic beam interacts with the target nuclei while the
nuclei are surrounded by a number of bound electrons. It leads to that 
the fusion cross section is increased.
At low and moderate energies, the fusion cross section of "bare" charged nuclei
colliding with the relative momenta $p$ and reduced mass $M$ (in electron
mass unit) is expressed as Ref.~\cite{Lang}:
\begin{equation}
\label{sigma}
\sigma(E) = \frac{S(E)}{E}e^{-2\pi\eta}
\end{equation}
where $S(E)$ is the so-called astrophysical factor which incorporates all
nuclear aspects of the process, $E$ is the collision energy  of nuclei 
and $\eta=MZ_1Z_2/a_ep$( the latter factor comes from $\psi_{coul}(0)$- the 
Coulomb wave function of the internuclear motion at the origin) and $a_e$ is
the hydrogen Bohr radius.

The screened cross section differs by a factor $\gamma(E)$ defined as:

\begin{equation}
\label{gamma}
\gamma(E)  \equiv \frac{\sigma_{sc}}{\sigma_{ba}} 
= \frac{|\psi_E(0)|^2}{|\psi_{coul}(0)|^2}
\end{equation}
where $\sigma_{sc}$ and $\sigma_{ba}$ are fusion cross sections both of 
the screened and "bare" nuclei, and $\psi_E(0)$ is the wave function of 
the internuclear motion  at the origin, taking into account the bound 
electron around.

In this Letter we present the first quantum mechanical calculation of screening
effect by bound electron in 
$$
{(^7Be,e)} + p \longrightarrow {(^8B,e)} +\gamma
$$
nuclear fusion. This reaction is of no importance for the Sun luminosity but it
is of great interest in Solar Neutrino Puzzle. An analysis of Solar Neutrino
Experiment Data (see Ref.~\cite{Fiorentini}) shows that there is no room for $^7Be$ neutrinos. At the
same time $^8B$ neutrinos are presented. Since $^8B$ nuclei result from
the reaction: 
\begin{equation}
\label{creation}
{^7Be} + p \longrightarrow {^8B} +\gamma, 
\end{equation}
it looks impossible to explain the existence of $^8B$
neutrinos and the absence of $^7Be$ neutrinos in the frame of the 
Standard Solar Model. 

As  one of the main consequences of electron screening effect applied to
all solar fusions could be decrease in the Sun core temperature. Increased
values of fusion cross sections could lead to the Sun cooling with the same
observable Solar luminosity. 
However, just cool Sun models do not solve the Solar Neutrino
Problem (see Ref.~\cite{Langacker}).

We show that bound electron essentially enhances the fusion rate in 
comparison with the reaction (\ref{creation}) rate. 
Physics of this phenomena
could be easily understood in the framework of united nucleus approach 
(see below).
"Exact" solution of our problem is obtained in adiabatic approach for three
particle problem. We compare our solution with the two relevant approaches
(united nucleus and static approaches, described below) which give upper and 
lower estimate for the screening effect, in the considered here
three-particle picture. 

A finite ${^7Be-p}$ interaction radius also is of great importance. A larger
radius results in higher tunneling probability into the internal region and
thus to a higher ${^7Be}(p,\gamma){^8B}$ cross section. Theoretical model
Ref.~\cite{Csoto} also predicts increased astrophysical $S_{17}(0)$-factor.  
All above mentioned effects result in the increased ${^8B}$ production rate 
and the latter could be a really alternative to 
${^7Be} + e  \longrightarrow {^7Li} + \nu_e$ reaction.

\section{Method of Calculation}

We treat the Coulomb problem for three particles in the framework of the
adiabatic representation. The basic two-center eigenfunctions are derived
from the Schr\"odinger equation for two nuclei with electric charges 
$Z_1$ and $Z_2$ (in electron charge unit)
($Z_1 >  Z_2$) fixed  on a distance $R$ and for an 
electron  around them. The Schr\"odinger equation 
is transformed  into a infinite system of equations with separated variables. 
Our approximation consists in that we use only one two-center eigenfunction
corresponding to the ground state of the system. There are some arguments
for this approach. 

At first, the high energy states corrections (at fixed $R$)
are of order of magnitude about the ratio of the electron mass to the proton 
mass.

Then, excited energy levels correspond to the less energy of the united 
nucleus.
It leads to their exponentially small contribution into the nuclear fusion
rate in comparison with the ground state of the system.

At least, only the ground state energy of the electron in the field of
${^7Be}$ and $p$ nuclei, called $1S\sigma$ therm (see, for example
Ref.~\cite{Gershtein}) has the correct asymptotic behaviour (see below).

The three particle wave function is presented as:
$$
\Psi(\vec{R},\vec{r})=\phi(\vec{R}) \cdot \psi(\vec{R},\vec{r}),
$$
where $\phi(\vec{R})$ is the wave function of two colliding nuclei and
$\psi(\vec{R},\vec{r})$ is the electron wave function, which depends on the
internuclear distance $\vec{R}$.

The electron energy eigenvalue $U_{nlm}(R)$ also depends on the internuclear
distance. We use the tabulated values of $U_{nlm}$ from Ref.~\cite{Ponomarev} 
for our purposes.

\begin{figure}
\center{\mbox{\epsfig{file=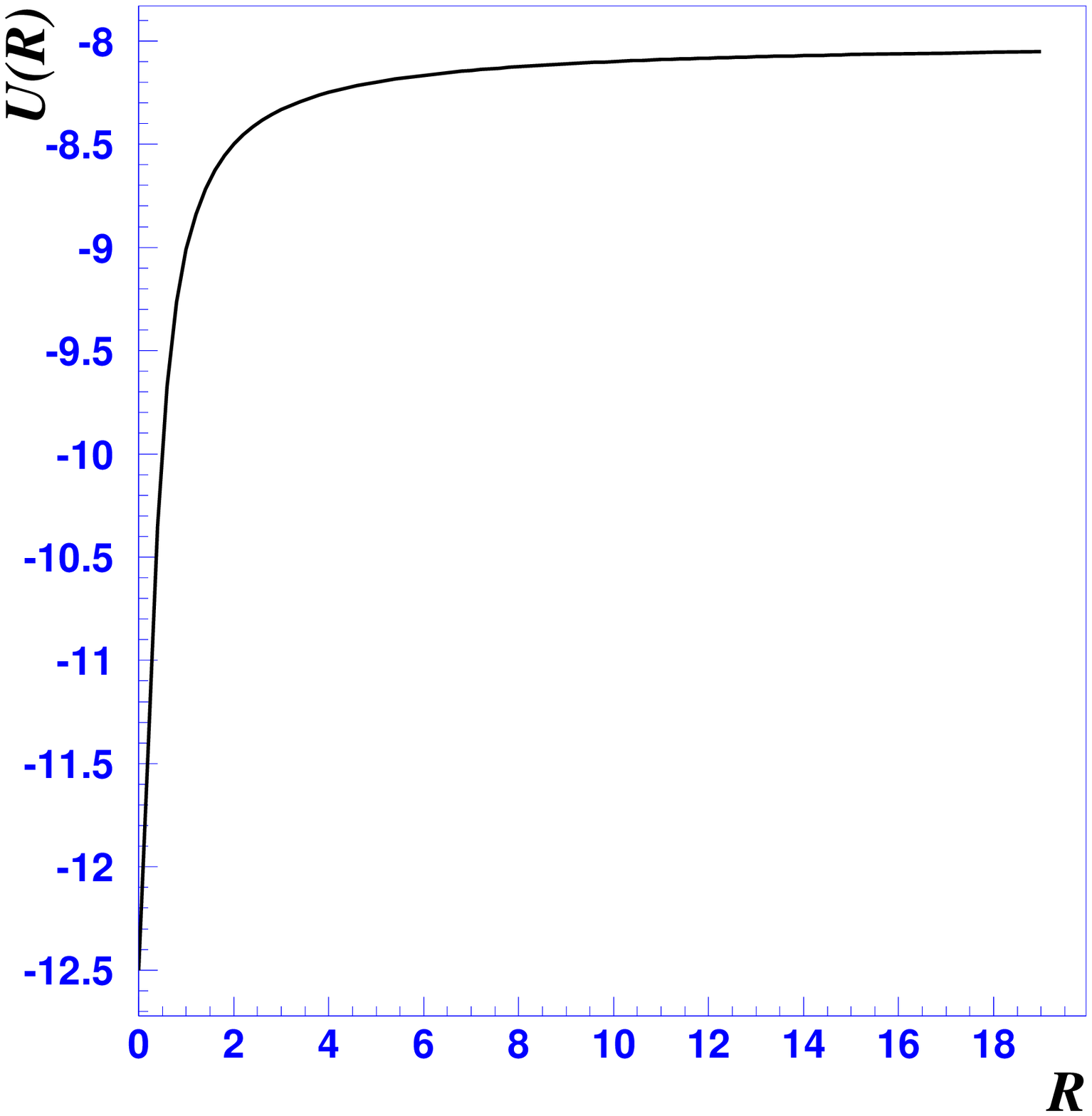}}}
        \protect\caption{Electron energy in the field of the two nuclei
                         $^7Be$ and $p$.
                         $R$ measured in Bohr radius, $U(R)$ given in 
                         $27.21$ eV unit }
\label{Beterm}
\end{figure}

In Fig.~\ref{Beterm} the values for the ground state are plotted. 
At $R \rightarrow 0$ the electron energy approaches to the energy of the united
ion: $U \rightarrow -\frac{(Z_1 + Z_2)^2}{2}$ (energy unit here is 27.21 eV )
and for  $R \rightarrow \infty$ the electron energy approaches to the energy 
of  the isolated ion $eZ_1$:  $U \rightarrow -\frac{Z_1^2}{2}$. 

We consider the case $Z_1 =4$ and $Z_2 = 1$.
In the adiabatic approach the electron energy $U_{nlm}(R)$ in the field of two
nuclei serves as an effective attraction potential for the nuclei. 
Evidently the main effect of the
electron screening comes from the collisions with zero orbital moment of
nuclei. The latter and the correct behaviour of $U_{nlm}(R)$ (at
$R\rightarrow 0 $ and $R\rightarrow \infty$) for the ground
state allow us to consider the case $n=l=m=0$. Let us denote $U_{000}$ as
$U(R)$. With the electron energy $U(R)$ in hand one can solve the scattering 
problem for two nuclei $^7Be$ and $p$. The total effective potential reads:
\begin{equation}
\label{poten}
V(R) = \frac{Z_1 Z_2}{R} + U(R)
\end{equation}
We calculate values of the
wave function of relative motion of the two nuclei at kinetic energies 
$0.1$ keV $\le E \le 100$ keV and compare the calculated $|\Psi(0)|^2$ with 
$|\Psi_{coul}(0)|^2 = \frac{2\pi\eta}{e^{2\pi\eta} - 1}$ - value of the
scattering wave function of two particles in the origin.

For simple estimations one can use well known united nucleus approach 
that consists in following. Consider a fusion of two 
nuclei from initial states with bound electrons. 
Let the  total negative energy of these bound electrons be $E_1 + E_2$. 
In the final state there is an united nucleus with bounded electrons around
and let the total energy of the electrons in this new ion be $E_{u}$. 
Thus the difference $\triangle E = E_1 + E_2 - E_{u}$ adds to the kinetic 
energy  of the two nuclei accelerating them. 
So, the united nucleus approach changes $|\Psi_E(0)|^2$ to
$|\Psi_{E + \triangle E}(0)|^2$. This is a rather good approach since the
main contribution to the value $|\Psi_E(0)|^2$ comes from internuclear
distances less then classical return point in the potential. 

Since united nucleus approach replaces decreasing $E(R)$ 
(see in Fig.~\ref{Beterm}) by the constant  $\triangle E$ this approach 
serves as an upper estimate.

Another approach uses static wave function of the bound electron. It means
that this wave function does not depend on the internuclear distance. 
Total three particle wave function is presented as a product:
$$
\Psi_s(\vec{R},\vec{r})=\phi_s(\vec{R})\cdot\psi_s(\vec{r}),
$$
where the wave function for the electron bounded on the nucleus with electric
charge $Z_1$ reads:
$
\psi_s(\vec{r})=\sqrt{\frac{Z_1^3}{\pi}}e^{-Z_1r},
$
and $\phi_s(\vec{R})$ is the wave function of two nuclei.
Averaging  the electron coordinates , $\phi_s(\vec{R})$ can be derived from
the  Schr\"odinger equation with the potential:
\begin{equation}
\label{static}
V_{st}(R) = 
\frac{Z_1Z_2}{R} - Z_2 \cdot 
\left(\frac{1-e^{-2RZ_1}}{R} - Z_1e^{-2RZ_1}\right).
\end{equation}
Kinetic energy of the colliding nuclei is counted from the electron
eigenvalue energy in the field of $Z_1$ nucleus.
This approach works well for nuclei with relative velocity $v$ much greater
then electron velocity $v_e$: $v >> v_e$. 

Attractive part of the potential
$V_{st}$ approaches to the value :$Z_1Z_2$ (in 27.21 eV unit)
at $R \longrightarrow 0$. Since $Z_1Z_2 < \frac{(Z_1+Z_2)^2}{2}$-the value
of the electron energy in united ion, the static approach
gives a lower estimate for an enhancement factor. 

We use all three approximations in order to compare their validity. 

\section{Results}

The numerical solution of the scattering problem of the two nuclei in the
potential (\ref{poten}) was obtained on the mesh on $R$ for $R=0$ and
$R=R_{max}$ in Numerov's scheme. 
We varied step $h$ and $R_{max}$ in order to ensure that final result 
is not changed substantially. 
Also for checking purposes we reproduced the value of $|\Psi_{coul}(0)|^2$ 
substituting $U(R) = 0$.

\begin{figure}
\center{\mbox{\epsfig{file=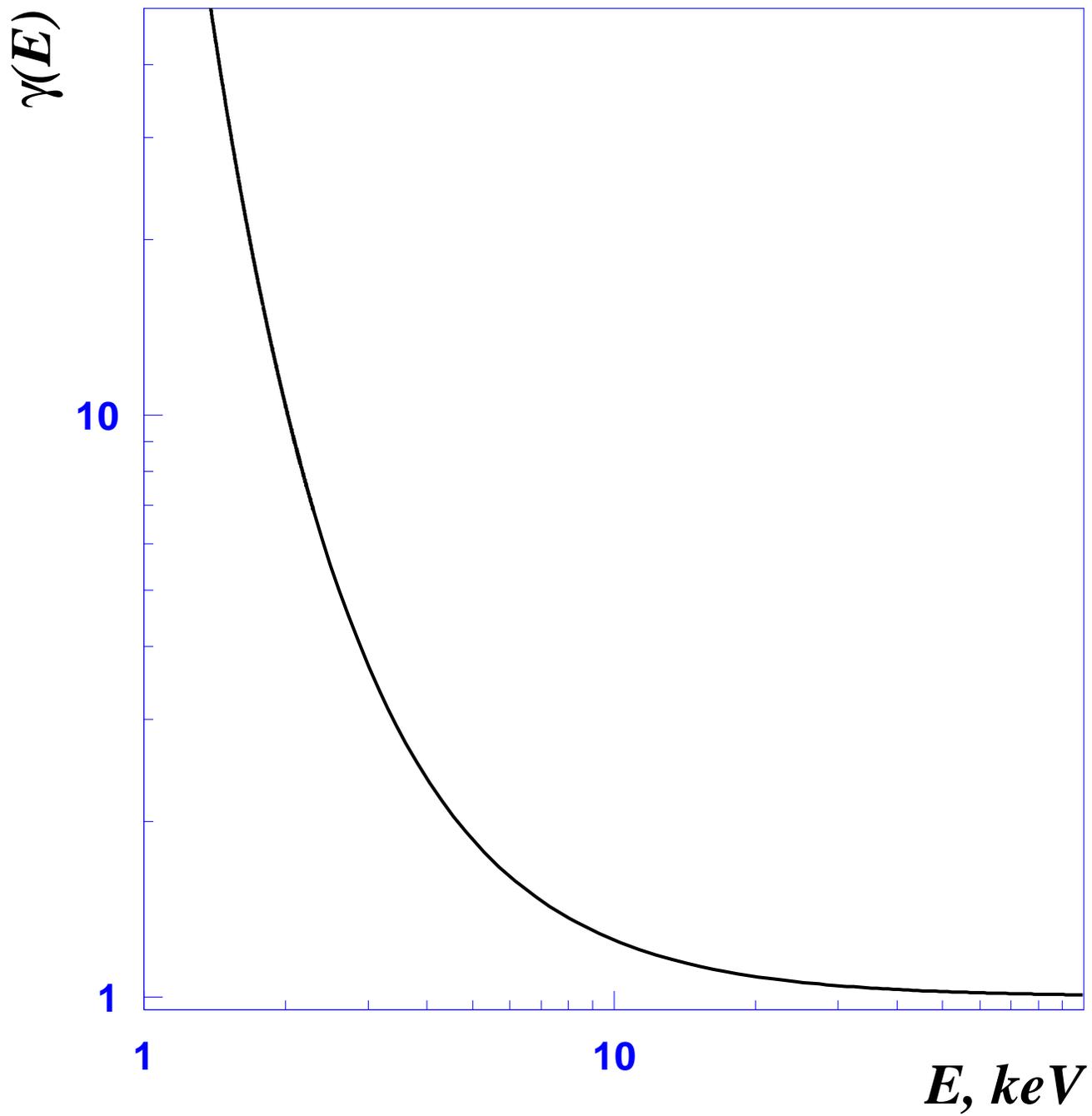}}}
        \protect\caption{Enhancement of nuclear fusion rate due to the
                         electron screening}
\label{enhan}
\end{figure}

\begin{figure}
\center{\mbox{\epsfig{file=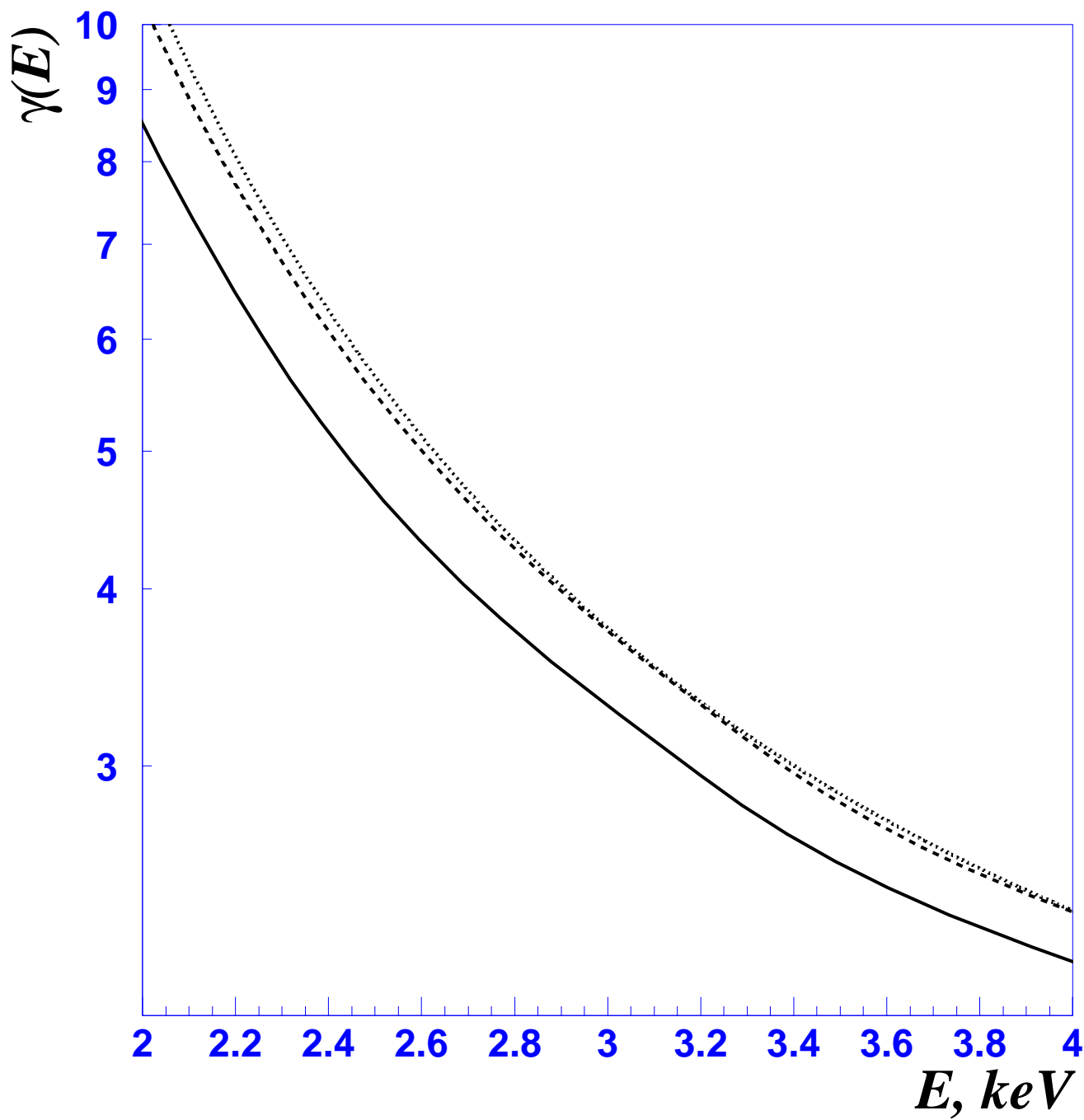}}}
        \protect\caption{$\gamma(E)$ factor for united nucleus (dotted line),
                         "exact" numerical solution (dashed line) and static
                           approaches (solid line) }
\label{gam}
\end{figure}

The enhancement factor (\ref{gamma}) is plotted in Fig.~\ref{enhan}. 
In Fig.~\ref{gam} we compare the "exact" numerical solution with both 
united nucleus and static approaches. 

At kinetic energies above $2$ keV a very good agreement between all three
approaches is obtained,
though at lower energies the united nucleus approach overestimates and the
static approach
underestimates an electron screening effect. 
It is easy to see that simple
united nucleus prescription gives much closer values for electron screening 
effect then the static approach. Thus the united nucleus approach being 
compared with "exact" numerical solution could be easily used in stellar
evolution code.

As one can see, electron screening could essentially increase nuclear fusion
rate and must be taken into account both for Earth experiments at low
energies and for an understanding of stellar phenomena. 

Deficit of $^7Be$
neutrinos in comparison with $^8B$ neutrinos (called "second Solar Neutrino
Problem", see Ref.~\cite{Bahcall}) could be a good example of this effect.
At the same Solar Model input data the flux of  $^8B$ neutrinos increases 
due to the increased rate of  $^8B$ nuclei production and the flux of $^7Be$ 
neutrinos decreases due to the alternative reaction (\ref{creation}).
At mean kinetic energy in the Sun core at temperature $T = 15 T_6$ 
($T_6 = 10^6$K) this effect is $15$ times of magnitude.

However due to the Gamov's "window" around $17-20$ keV the screening effect
of the bound electron is reduced to $10\%$. But if some nuclear fusion
reaction go faster in the Sun core (due to the electron screening) then the
temperature in the Sun core could be lower. Under this assumption it might
be possible to explain the Solar Neutrino Problem. 

In conclusion, we have performed quantum mechanical calculation of the
screening effect taking place in  
${(^7Be,e)} + p \longrightarrow {(^8B,e)} +\gamma$ reaction at finite
kinetic energy of the nuclei. We compared our results with two approaches
giving upper and lower limits for the screening effect and studied the validity
of the approaches. 
Possible application to the Solar Neutrinos is briefly discussed.

\newpage

\end{document}